\documentclass[12pt]{article}
\usepackage{hyperref,amsfonts,amssymb, amsmath,mathrsfs}
\usepackage[english]{babel}

\def\nn{\nonumber }
\def\bq{ \begin{equation} }
\def\eq{ \end{equation} }
\def\ben{ \begin{eqnarray} }
\def\en{ \end{eqnarray} }

\def\on#1#2{\mathop{\vbox{\ialign{##\crcr\noalign{\kern2pt}
$\scriptstyle{#2}$\crcr\noalign{\kern2pt\nointerlineskip}
\kern-2pt$\hfil\displaystyle{#1}\hfil$\crcr}}}\limits}
\newtheorem{prop}{Theorem}

\begin{document}


\title{On integrable system on $S^2$ with the second integral quartic in the momenta.}
\author{
 A.V. Tsiganov\\
\\
\it\small St.Petersburg State University, St.Petersburg,
Russia}

\date{}
\maketitle

\begin{abstract}
We consider integrable system on the sphere $S^2$ with an
additional integral of fourth order in the momenta. At the
special values of parameters this system coincides with the
Kowalevski-Goryachev-Chaplygin system.
\end{abstract}

\section{Introduction}
Let us consider particle moving on the sphere $S^{2}=\{x\in
\mathbb R^3, |x|=a\}$. Entries of the vector $x$ and angular
momentum vector $J=p\times x$ are coordinates on the phase
space $T^*S^2$ with the following Poisson brackets
\begin{equation}\label{e3}
\,\qquad \bigl\{J_i\,,J_j\,\bigr\}=\varepsilon_{ijk}J_k\,,
\qquad \bigl\{J_i\,,x_j\,\bigr\}=\varepsilon_{ijk}x_k \,,
\qquad \bigl\{x_i\,,x_j\,\bigr\}=0\,,
\end{equation}
where $\varepsilon_{ijk}$ is the totally skew-symmetric
tensor. The Casimir functions of the brackets (\ref{e3})
\bq\label{caz}
A=\sum_{i=1}^3 x_i^2=a^2,\qquad B=\sum_{i=1}^3 x_iJ_i=0\,,
\eq
are in the involution with any function on $T^*S^2$. The
phase space $T^*S^2$ is four dimensional symplectic manifold.
So, for the Liouville integrability of the corresponding
equations of motion it is enough to find two functionally
independent integrals of motion.

In this note we discuss an integrable system on $T^*S^2$
possessing integrals of second and fourth order in the
momenta $J_k$. The corresponding Hamilton function has a
natural form, i.e. it is a sum of a positive-definite kinetic
energy and a potential. So, according to Maupertuis's
principle, this natural integrable system on $T^*S^2$
immediately gives a family of integrable geodesic on $S^2$.
Integrals of the geodesic are also polynomials of the second
and fourth degrees.

Remind that description of all the natural Hamiltonian
systems on closed surfaces admitting integrals polynomial in
momenta is a classical problem \cite{dar94}. For the systems
with polynomial in momenta integrals of degree one or two
there exists a complete description and classification
\cite{fom00}.

The Kowalevski top is an example of a conservative system on
$S^2$ which possesses an integral of degree four in momenta
\cite{kow89}. Later Goryachev \cite{gor16} and Chaplygin
\cite{ch03} found generalization of the Kowalevski system on
$S^2$. Recently, these results were extended in \cite{sel99}.

The main aim of this note is to consider some another
generalization of the Kowalevski-Goryachev-Chaplygin system
using the reflection equation theory \cite{skl88}.

\section{Generic case}
Following to \cite{ts04} let us consider Lax matrix for the
generalized Lagrange system
\[
T(\lambda)=\left(%
\begin{array}{cc}
 A & B \\
 B^* & D \\
\end{array}%
\right)(\lambda)
\]
with the following entries polynomial in the spectral
parameter $\lambda$
\ben
A(\lambda)&=&\lambda^2-2\lambda\alpha J_3
+\Bigl(\alpha^2-f(x_3)\Bigr)J_3^2-J_1^2-J_2^2-g(x_3)\,,\nn\\
\nn\\
B(\lambda)&=&(x_1+ix_2)m(x_3)\lambda+J_3(x_1+ix_2)\ell(x_3)+(J_1+i
J_2) n(x_3)\,,
\label{ABC}\\
\nn\\
 D(\lambda)&=& -n(x_3)^2\nn\,.
\en
Here $\alpha$ is an arbitrary numerical parameter, $f,g,m,n$
and $\ell$ are some functions of $x_3$ and of the single
non-trivial Casimir $a=\sqrt{x_1^2+x_2^2+x_3^2}$ (\ref{caz}).

The trace of the matrix $T(\lambda)$
\[\mbox{\rm tr}\,T(\lambda)=A(\lambda)+D(\lambda)=\lambda^2-\lambda H_L+K_L\]
gives rise integrals of motion in the involution for the
generalized Lagrange system
\[
H_L=2\alpha J_3\,,\qquad K_L=
\Bigl(\alpha^2-f(x_3)\Bigr)J_3^2-J_1^2-J_2^2-g(x_3)-n(x_3)^2\,.
\]
The corresponding equations of motion may be rewritten in the
form of the Lax triad
\[
\dfrac{d}{dt}T(\lambda)=\Bigl[T(\lambda),M(\lambda)\Bigr]+N(\lambda),\qquad
\mbox{\rm tr}\,N(\lambda)=0\,.
\]

With algebraic point of view coefficients of the trace of
 $T(\lambda)$ give rise commutative subalgebra in the complete
Poisson  algebra generated by entries $T_{ij}(\lambda)$. All
the generators of this subalgebra are linear polynomials on
coefficients of entries $T_{ij}(\lambda)$, which are
interpreted as integrals of motion for integrable system
associated with matrix $T(\lambda)$.

Some special commutative subalgebras generated by
quad\-ra\-tic polynomials on coefficients of
$T_{ij}(\lambda)$ were considered in \cite{ts04}. These
subalgebras were associated with five integrable systems on
$S^2$ with an additional integral of motion third order in
the momenta.

According to \cite{skl88,sokts2}, we can try to construct
another commutative subalgebra generated by quad\-ra\-tic
polynomials on coefficients of $T_{ij}(\lambda)$, which are
integrals of motion for another integrable system associated
with the same matrix $T(\lambda)$.

Namely, using matrix $T(\lambda)$ (\ref{ABC}) and standard
machinery of the reflection equation theory
\cite{skl88,sokts2} we can construct another $2\times 2$
matrix
\bq
L(\lambda)={\mathcal K}_-(\lambda)\,T(\lambda)\,{{\mathcal
K}}_+(\lambda)\,\left(\begin{array}{cc}
 0 & 1 \\
 -1 & 0
\end{array}\right)T^t(-\lambda)\left(\begin{array}{cc}
 0 & 1 \\
 -1 & 0
\end{array}\right)\,,\label{Laxref}
\eq
with the following trace
\bq\label{P3}
\mbox{\rm tr}\, L(\lambda)=
-\lambda^6-F_1\,\lambda^4\,-F_2\lambda^2-F_3\,.
\eq
Here the superscript $t$ stands for matrix transposition and
matrices ${\mathcal K}_\pm(\lambda)$ are numerical solutions
of the reflection equation associated with the $r$-matrix of
$XXX$ type.

Let us begin with the following partial numerical solutions
the reflection equation
\bq \label{kpm}
\mathcal K_+(\lambda)=\left(%
\begin{array}{cc}
 b_1\lambda+b_0 & \lambda \\
 0 & -b_1\lambda+b_0
\end{array}%
\right),\qquad \mathcal K_-(\lambda)=\mathcal
K_+^t(\lambda)\,,
\eq
which depend on two arbitrary parameters $b_0$ and $b_1$
only.

Substituting $T(\lambda)$ (\ref{ABC}) and $\mathcal K_\pm$
(\ref{kpm}) into $L(\lambda)$ (\ref{Laxref}) one gets that
function $F_3$ in (\ref{P3}) depends of variables $x_3$ and
$J_3$ only. So, if we want to consider integrable system
differed from the generalized Lagrange system we have to put
$F_3=const$. It leads to the following expressions of the
functions $f(x_3)$ and $g(x_3)$
\bq\label{fg}
f(x_3)=
\alpha^2-\dfrac{2\ell(x_3)x_3}{n(x_3)}+\dfrac{\ell(x_3)^2(a^2-x_3^2)}{n(x_3)^2},\qquad
g(x_3)=\dfrac{d}{n(x_3)^2}\,,
\eq
where $d$ is arbitrary numerical parameter.
\begin{prop}
If functions $f(x_3)$ and $g(x_3)$ are given by (\ref{fg})
then the third coefficient $F_3$ in (\ref{P3}) is a constant
\[F_3=2b_0^2d,\]
while two remaining coefficients $F_1$ and $F_2$ are in the
involution on $T^*S^2$ if and only if
\[
n(x_3,a)=
c_1\sin\left(\alpha\arctan\left(\dfrac{x_3}{\sqrt{a^2-x_3^2}}\right)\right)
 +c_2\cos\left(\alpha\arctan\left(\dfrac{x_3}{\sqrt{a^2-x_3}}\right)\right).
\]
Here $\alpha,c_1,c_2$ are arbitrary parameters and all
another functions in (\ref{ABC}) are equal to
\[
\begin{array}{lll}
 \alpha=0,\qquad & m=0,\quad &
 \ell=n\dfrac{\sqrt{x_3^2-a^2}-x_3(\ln(x_3+\sqrt{x_3^2-a^2})+c_3)}{
 (x_3^2-a^2)(\ln(x_3+\sqrt{x_3^2-a^2})+c_3)},\\
 \\
 \\
 \alpha\neq 0 & m=-\dfrac{n'}{\alpha},\qquad & \ell=\dfrac{(\alpha^2n-x_3n')n}{(x_3^2-a^2)n'},
\end{array}
\]
where $n'=\dfrac{\partial n(x_3,a)}{\partial x_3}$.
\end{prop}
The proof is straightforward.

So,  two functions $F_1$ and $F_2$ are in the involution
$\{F_1,F_2\}=0$  on the phase space $T^*S^2$. Moreover,
direct calculation yields that they are functionally
independent functions on $T^*S^2$. It means that these
functions $F_1$ and $F_2$  define an integrable system on the
sphere.

Integrals of motion $F_1$ and $F_2$ are quadratic and quartic
polynomials in the momenta. For instance, at $\alpha\neq 0$
the corresponding Hamilton function is equal to
\ben
H=\dfrac{F_1}{2}&=&J_1^2+J_2^2+\left(2\alpha^2+
(a^2-x_3^2)\dfrac{\ell^2}{n^2}-2x_3\dfrac{\ell}{n}\right)J_3^2+2b_0\dfrac{n'}{\alpha}\,x_1\nn\\
\label{H-gen}\\
&+&2b_1\bigl(J_1n+J_3x_1(\ell-2n')\Bigr)
+b_1^2\left(n^2+\dfrac{(a^2-x_3^2)n'^2}{\alpha^2}\right)+\dfrac{d}{n^2}\,.\nn
\en
For brevity we do not present the second integral of motion
$F_2$ explicitly. This function $F_2$ may be restored from
the definitions (\ref{ABC}-\ref{P3}) and conditions of the
Theorem 1.

If we consider more generic solutions of the reflection
equations
\[
\mathcal K_+(\lambda)=\left(%
\begin{array}{cc}
 b_1\lambda+b_0 & \lambda \\
 0 & -b_1\lambda+b_0
\end{array}%
\right),\qquad \mathcal K_-(\lambda)=\left(%
\begin{array}{cc}
 d_1\lambda+d_0 & 0\\
 \lambda & -d_1\lambda+d_0
\end{array}%
\right)\,,
\]
which depend of four parameters,  one gets the same integrals
of motion up to rescaling of $x$ and rotations
\bq\label{c-trans}
x\to b\, U\, x\,,\qquad J\to U J\,,\qquad U=\left(%
\begin{array}{ccc}
 1 & 0 & 0 \\
 0 & \cos(\phi) & -\sin(\phi) \\
 0 & \sin(\phi) & \cos(\phi)
\end{array}%
\right)\,,
\eq
where $b$ and $\phi$ are the suitable parameters.

Up to such transformations integrals of motion $F_1$ and
$F_2$ depend of five numerical parameters $\alpha$, $b_0$,
$b_1$, $c_1/c_2$ and $d$. Remind, that another two parametric
family of integrable systems on the sphere with fourth order
integral of motion was studied in \cite{sel99}. However,
ansatz for the Hamilton function proposed in \cite{sel99}
looks like more restrictive than Hamiltonians  (\ref{H-gen}).
The relations between these systems will be studied in the
forthcoming publications.

\section{Special cases}

\par
At $\alpha=0$ the Hamiltonian reads as
\ben
H_0=\dfrac{F_1}{2}&=&J_1^2+J_2^2+\left(\dfrac{x_3^2}{x_3^2-a^2}-\dfrac{1}{(\ln(x_3+\sqrt{x_3^2-a^2})+c_3)^2}\right)J_3^2\nn\\
\label{H-0}\\
&+&2b_1c_2\left(J_1
-\dfrac{(\ln(x_3+\sqrt{x_3^2-a^2})+c_3)x_3-\sqrt{x_3^2-a^2}}{(x_3^2-a^2)(\ln(x_3+\sqrt{x_3^2-a^2})+c_3)}\,x_1J_3\right)\,.
\nn
\en
It define new integrable system on the sphere, which depend
of two parameters $b_1c_2$ and $c_3$ only.

If $\alpha= 1$ or $\alpha=2$ one gets
\[
n = c_1x_3+ c_2\sqrt{a^2-x_3^2}\, \qquad\mbox{\rm and}\qquad
n(x_3)=c_1x_3\sqrt{a^2-x_3^2}+c_2(a^2-2x_3^2)\,,\]
respectively.  Even in these particular cases the
corresponding Hamiltonian $H$ (\ref{H-gen}) remains a huge
function. We will present it imposing some additional
restrictions only.

At $\alpha=1$ and $c_2=0$ the Hamiltonian (\ref{H-gen}) is
equal to
\bq\label{H1}
H_1=\left.\dfrac{F_1}2\right|_{c_2=0}=J_1^2+J_2^2+2J_3^2+
2b_1c_1(J_1x_3-2J_3x_1)+2b_0c_1x_1+\dfrac{d}{c_1^2x_3^2}
-b_1^2c_1^2a^2\,.
\eq
After canonical transformation
\[
J_1\to J_1 - c_1b_1x_3,\quad J_2\to J_2, \quad J_3\to
J_3+c_1b_1x_1,\qquad x_k\to x_k
\]
this Hamiltonian $H_1$ (\ref{H1}) reads as
\bq\label{H-Kow}
H_1=J_1^2+J_2^2+2J_3^2+2c_1b_0x_1+c_1^2b_1^2(x_1^2-x_2^2)+\dfrac{d}{c_1^2x_3^2}\,.
\eq
It is the Hamilton function for the
{K}owalewski-{C}haplygin-{G}oryaschev top.  Canonical
transformations (\ref{c-trans})  allow us to rewrite the
Hamilton function (\ref{H-Kow}) for the
Kowalevski-Goryachev-Chaplygin system in the standard form
\cite{kuzts1}. The corresponding Lax matrix $L(\lambda)$
(\ref{Laxref}) was constructed in \cite{kuzts1}. The
separation of variables associated with this matrix is
discussed in \cite{ts02b}.

At $\alpha=1$ and $c_1=0$ the corresponding Hamilton function
(\ref{H-gen}) is equal to
\ben
\widetilde{H}_1=\left.\dfrac{F_1}2\right|_{c_1=0}&=&
J_1^2+J_2^2+\dfrac{2x_3^4-a^4}{x_3^2(x_3^2-a^2)}\,J_3^2+2c_2b_0\dfrac{x_1x_3}{\sqrt{x_3^2-a^2}}
+\dfrac{d}{c_2^2(x_3^2-a^2)}\nn\\
\nn\\
\label{tH1}\\
&+&
2c_2b_1\left(\sqrt{x_3^2-a^2}J_1-\dfrac{2x_3^2+a^2}{x_3\sqrt{x_3^2-a^2}}\,x_1J_3\right)
-b_1^2c_2^2a^2\,.\nn
\en

At $\alpha=2$ and $c_1=0$  it has the form
\ben
\widetilde{H}_2=\left.\dfrac{F_1}2\right|_{c_1=0}&=&J_1^2+J_2^2+\left(5+\dfrac{a^2}{x_3^2}\right)J_3^2
-4b_0c_2x_3x_1+\dfrac{d}{c_2^2(2x_3^2-a^2)^2}\nn\\
\nn\\ \nn\\
&-&2c_2b_1\left(
(2x_3^2-a^2)J_1-\dfrac{6x_3^2+a^2}{x_3}\,x_1J_3 \right)
+b_1^2c_2^2a^4\,.\nn
\en
These Hamiltonians define new integrable systems on the
sphere, which depend of three arbitrary parameters only.

\section{Summary}
Using the Lax matrix for the generalized Lagrange system and
the standard construction of the commutative subalgebras from
the reflection equation theory we construct new integrable
system on the sphere. The corresponding Hamilton function is
given by (\ref{H-0}, \ref{H-gen}) while the second integral
is fourth order polynomial in the momenta.

These integrals depend of five numerical parameters
$\alpha,b_0,b_1,c_1/c_2$ and $d$ up to canonical
transformations. At the special values of parameters we
recover the Kowalevski-Goryachev-Chaplygin system.

The research was partially supported by RFBR grant
02-01-00888.

\end{document}